\documentclass[reqno]{amsart}
\usepackage{graphics,amsmath}

\def\d{{\rm d}} 

\newcommand{\be}{\begin{equation}}
\newcommand{\ee}[1]{\label{#1} \end{equation}}
\newcommand{\ba}{\begin{eqnarray}}
\newcommand{\ea}[1]{\label{#1} \end{eqnarray}}
\newcommand{\nl}{\nonumber \\}
\newcommand{\re}[1]{(\ref{#1})}

\def\re#1{(\ref{#1})}   

\begin{document}

\title{Thermodynamics and flow-frames for dissipative relativistic fluids}

\author{P. V\'an$^{1,2,3}$ and T.S. Bir\'o$^{1}$} 
\address{ 
    $^1$Dept. of Theoretical Physics, Wigner Research Centre for Physics, Institute for Particle and Nuclear Physics, \  
      H-1525 Budapest, Konkoly Thege Mikl\'os \'ut 29-33, Hungary; \\
    $^2$Dept. of Energy Engineering, Budapest Univ. of Technology and Economics,\\
      H-1111, Budapest, M\H uegyetem rkp. 3-9,  Hungary; \\
    $^3$Montavid Thermodynamic Research Group
} 

\date{\today}

\begin{abstract}

A general thermodynamic treatment of dissipative relativistic fluids is introduced, 
where the temperature four vector is not parallel to the velocity field of the fluid. 
Generic stability and kinetic equiblirium points out a particular thermodynamics, 
where the temperature vector is parallel to the enthalpy flow vector and 
the choice of the flow fixes the constitutive functions for viscous stress and heat. 
The linear stability of the homogeneous equilibrium is proved in a mixed particle-energy flow-frame.
\end{abstract}

\maketitle

\section{Introduction}

It is commonly accepted that in relativistic fluids there is a freedom in the definition of the velocity field \(u^{a}\), in the sense that one may fix it to different physical four vectors. The well known common choices are when the flow is defined by the particles (this is the Eckart flow-frame \cite{Eck40a3}), or by the energy (this is the Landau-Lifshitz flow-frame \cite{LanLif59b}). Therefore the velocity field of a one component fluid is a somewhat vaguely defined physical property, belonging to the flow of volatile quantities, once the energy, once the conserved charge. That attitude assumes that the physics is invariant when changing the flow-frame. 

Other evidently flow-frame dependent properties of a relativistic dissipative fluid theory are related to the thermodynamic framework. Gibbs relation is classically given in the following form \cite{Eck40a3,GroAta80b}:
\be
ds = \beta de - \alpha dn,
\ee{Gibbs_r}

where $s=u_aS^a$ is the entropy density, $e=u_au_bT^{ab}$ is the energy density, $n=u_aN^a$ is the particle number density, $\beta$ is the reciprocal temperature and $\alpha$ is the chemical potential divided by the temperature. This non-relativistically motivated formula is defined by the parts of the covariant $S^a$ and $N ^a$ entropy and particle number density vectors and the $T^{ab}$ energy-momentum tensor, that are parallel to the $u^a$ velocity field of the fluid. It is not evident that this particular form, or any generalization of that is invariant when changing the flow-frame. 

The flow-frame dependence of the different relativistic fluid theories is more apparent
when focusing on generic stability, on one of the fundamental problems in relativistic fluids. Thermodynamics is related to stability \cite{Mat05b}. One expects in any reasonable theory of continua that the homogeneous equilibrium of an insulated system is asymptotically stable and the conditions of this stability are ensured by thermodynamics. In particular for one component fluids the concave entropy -- thermodynamic stability -- and nonegative entropy production -- nonnegative transport coefficients -- should be the only conditions. A weaker requirement is the linear stability of homogeneous equilibrium, that we call generic stability. However, the first order Eckart theory is violently unstable in the Eckart flow-frame and stable in the Landau-Lifshitz one. Therefore, according to the usual explanations,
we must consider a higher order theory. For example the Israel-Stewart theory is believed to be a stable one. 

However, the generic stability is proved only in the Eckart flow-frame and the conditions given by Hiscock and Lindblom are very complicated \cite{HisLin83a,HisLin87a} and lacking a simple physical interpretation. Other higher order theories and other flow-frames are not analysed from this point of view.

Our final example of flow-frame dependence comes from kinetic theory. There one assumes a temperature four vector, which is parallel to the flow. Then an ideal fluid is defined by the following form of particle number density and energy-momentum:
\be
N^a_0 = n u^a, \qquad \text{and} \qquad T_0^{ab} = eu^au^b + p \Delta^{ab}.
\ee
where $p$ is the static pressure. However, a simple flow-frame transformation leads to
\be
N_0^a = n_0 u^a + j_0^a,  \qquad \text{and} \qquad 
T^{ab}_0 = e_0 u^a u^b +q_0^a u^b + u^a q_0^b + P^{ab}_0.
\ee{kEM}

Here $j_0^a = n_0 w^a$, $q_0^a = h_0 w^a$ and $P_0^{ab} = p \Delta^{ab} + h w^a w^b$ and $w^a$ is related to the u-orthogonal part of the temperature vector:
\be
\beta^a = \frac{u^a+w^a}{T}.
\ee{temp}

Therefore the above formal concept of ideal fluid is flow-frame dependent.

According to our investigations the generalization of thermodynamics, where the temperature vector is not parallel to the four velocity of the fluid improves the stability properties \cite{VanBir08a,Van09a,BirVan10a,VanBir12a}. The analysis of the problematics of flow-frames resulted in the following form of the entropy production \cite{VanBir13p}:
\ba
\Sigma &=& \partial_a S^a = (nw^a-j^a) \partial_a\alpha + (q^a-hw^a)(\partial_a \beta+\beta \dot u_a) 
\nonumber\\
&+& (\Pi^{ab}-q^{(a}w^{b)})
\partial_a\beta_b+ q^{[b}w^{a]}\partial_a(\beta (u_b-w_b))\geq 0. 
\ea{etrpr}

Here the background Gibbs relation was assumed in a form that corresponds to the general form of the temperature \re{temp}:
\be
ds = \beta de + \beta w_a dq^a - \alpha dn.
\ee{Gibbs_g}

We may have an important observation here. The classical assumption $w^a =0$, 
which may be considered as an equation of state 
specifying a flow-frame fixed to the thermometer, leads to the Eckart form of the entropy production. The corresponding fluid theory is unstable \cite{HisLin85a}. 
There are two particular choices that lead to generic stable relativistic hydrodynamics. 
The first one introduces an entropy density that depends on the ''length''
of the energy-momentum vector: $s=s(E,n)$, where $E=\sqrt{u_bT^{ab} T_{ac}u^c}$. 
In this case $w^a = q^a/e$ \cite{Van08a}. 
However, this result is not quite compatible with the concept of kinetic equilibrium. 
The second one introduces a different thermodynamics, where the 
Gibbs relation is defined according to kinetic equilibrium: $w^a = q^q/h$, 
with $h=e+p$ being the enthalpy density \cite{Van11p}.  
The hydrodynamics based on this suggestion is proved to be generic 
stable in an Eckart frame \cite{VanBir12a}.  

In these mentioned investigations we did not consider the compatibility of thermodynamic requirements and frame fixing. 
In the present paper we develope a general thermodynamic analysis based on the kinetic compatible 
Gibbs relation. In the derived expression of entropy production one can identify two thermodynamic force-current pairs for the three constitutive quantities: the diffusion current density, heat current and viscous stress. 
In this way a flow-frame fixes the constitutive framework, but does not introduce any restriction on the physics. 

In this work we do not investigate the question of causality, which is traditionally considered the most important problem in relativistic context and the main motivation of the development of second order fluid theories \cite{Mul69a}. There is also a general belief that causality implies stability \cite{DenEta08a,PuEta10a}. One may have two comments in this respect. First of all the continuum limit reduces the propagation spead of real signals \cite{Fic92a,KosLiu00a,Cim04a} and, on the other hand, several second order theories are not proved to be symmetric hyperbolic or divergence type \cite{LiuAta86a,PerCal10a,Ger01m,GerLin90a}. An important benchmark is the comparison of 
the numerical solutions of different fluid theories and the Boltzmann equation in various cases \cite{Mol09a,BouEta10a}.

This paper is organized as follows. In the next section we introduce the kinetic compatible relativistic thermodynamics. In the third section the entropy production is calculated and analysed. In the fourth section the generic stability of the dissipative hydrodynamics is proved in a mixed particle-energy flow-frame.

\section{Thermodynamics of motion}

In this paper we use the Lorentz form  with the sign convention
($+$,$-$,$-$,$-$), the indexes 
$a,b,c,...$ run over $0, 1, 2, 3$. We use natural units, $c = 1$. 

For isotropic media the entropy density \(S^a\) is connected to the particle number density 
\(N^a\) and energy-momentum density \(T^{ab}\) by the isotropic pressure:
\be
   p\beta^a = S^a+\alpha N^a-\beta_bT^{ab}.
\ee{eqrel}

Standard kinetic theory definitions and calculations for ideal substances satisfy the above expressions in thermodynamic equilibrium \cite{Isr63a,Van11p}.  
Eq. \re{eqrel} is our starting point in the following investigations, we 
propone its validity for nonideal substances and also out of hydrodynamic equilibrium. 
An arbitrary flow, $u^a$, decomposes the quantites of the above expression as follows:
\ba
S^{a}   &=& su^a+J^a, \label{S_sp}\\ 
N^{a}   &=& nu^a+j^a, \label{N_sp}\\ 
T^{ab}  &=& eu^au^b+q^au^b+u^aq^b+P^{ab},\label{T_sp}\\
\beta^a &=& \beta(u^a+w^a).
\ea{b_sp}

Here \(n\) is the flow-frame particle number density, \(j^{a}\) is the diffusion current, the non-convective particle number current density in this frame, \(e\) is the energy density, \(q^{a}\) is the momentum density and the heat current and \(P^{ab}\) is the pressure tensor. These components are flow-frame dependent, in particular \(j^{a}u_a=0\), \(q^au_a=0\) and \(P^{ab}u_b=0\). Then the u-timelike and u-spacelike parts of \re{eqrel} 
deliver
\ba
s=\beta h+\beta w_bq^b -\alpha n, \label{entr_d}\\
J^a= \beta(q^a+w_b \Pi^{ab} -\mu j^a).
\ea{entr_c}
Here we have introduced the viscous stress, as $\Pi^{ab} = P^{ab} + \Delta^{ab}p$.

The covariant analysis of kinetic equilibrium  requirements resulted in the following Gibbs relation \cite{VanBir12a}:
\be
ds = \beta_a \d E^a +\beta p w_a \d u^a -\alpha \d n. 
\ee{Gibbs_m}
Then one can obtain the corresponding Gibb-Duhem relation with the help of \re{entr_d}:
\be
\beta \d p + h \d\beta +q^a \d \beta_a - \beta w^a h \d u_a - n \d\alpha  =0. 
\ee{GD}

\section{Flow-frame independent entropy production}

Introducing the substantial time derivative $\frac{d}{d\tau}:= u^a\partial_a$,
abbreviated by and overdot, the balances for the particle current density and energy-momentum 
tensor are expressed in terms of the local rest frame quantities:
\ba
\partial_a N^a &=& \dot n + n\partial_au^a + \partial_aj^a=0,\label{pnumc_bal}\\
\partial_b T^{ab} &=&  \dot e u^a +  e\dot u^a + eu^a\partial_bu^b +\dot q^a 
    +q^a\partial_b u^b +    \nl
        &&     u^a\partial_b q^b+q^b\partial_b u^a +\partial_bP^{ab}=0^{a}. 
\ea{emomc_bal}

The energy and momentum balances are the time and spacelike components of the 
energy-momentum tensor balance projected onto the flow-frame: 
\ba
u_a\partial_b T^{ab} &=&  \dot e  +e\partial_bu^b +u_a\dot q^a +\partial_b q^b -P^{ab} \partial_bu_{a}= 0, \label{e_bal}\\
\Delta^a_c\partial_b T^{cb} &=&   e\dot u^a +\Delta^a_b\dot q^b +q^a\partial_b u^b + q^b\partial_b u^a +\Delta^a_c\partial_bP^{cb}  \nonumber\\
 &=& h\dot u^a +\Delta^a_b\dot q^b +q^a\partial_b u^b + q^b\partial_b u^a +\Delta^a_c\partial_b\Pi^{cb} - 
    \Delta^{ab} \partial_b p=0^a.
\ea{momc_bal}

The frame related quantities are important in the separation of the ideal and dissipative parts of the basic fields. This separation is best performed by analyzing the thermodynamical relations. In order to achieve this one postulates the existence of an additional vector field, the entropy current as a function of the basic fields \(S^{a}(N^a, T^{ab})\). It must not decrease by fulfilling the condition of the balances \re{pnumc_bal} and \re{emomc_bal}. That conditional   inequality can be best represented by introducing the Lagrange-Farkas multipliers\footnote{Lagrange multipliers are introduced for conditional extrema.  For conditional inequalities Gyula Farkas suggested analogous quantities and proved the corresponding theorem of linear algebra, called Farkas' lemma  \cite{Far894a,Min896b,Sch98b}.} \(\alpha\) and \(\beta^a\), respectively:\ 
\be
\Sigma := \partial_aS^a + \alpha \partial_aN^a - \beta_b\partial_aT^{ba} \geq 0.
\ee{entrcond}
The left hand side of this inequality shows, that the definition of the entropy production precedes the specification of the flow-frame. 

The separation of ideal and dissipative parts of basic physical quantities is a consequence to the choice of that flow-frame. Citing the authors of \cite{TsuKun12a}, when arguing about the uniqueness of the Landau-Lifshitz frame ``The uniqueness of the energy frame comes from ... the physical assumption that the dissipative effect comes from only the spatial inhomogeneity.``. 
However, the thermodynamic choice is primary,
the Landau-Lifshitz frame is an additional condition which may or may not be compatible with 
the preferred thermodynamics. 

Entropy is not an independent physical quantity: both the entropy density and the current density 
of the entropy are functions of the primary fields, as it is reflected by the Gibbs relation \re{Gibbs_m} 
and by \re{entr_c}. 
Moreover, the entropy inequality is conditional, the conservation of particle number, 
energy and momentum being the constraints. 
The best way to introduce these restrictions is by substituting the corresponding proper time derivatives:
\begin{gather}
 \dot s + s \partial_au^a +\partial_aJ^a = \nonumber\\
   \beta \dot e - \alpha \dot n + \beta w^a\dot q^a + 
   \beta(h w^a-q^a)\dot u^a +  s \partial_au^a +
  \partial_a\left[\beta q^a+\beta w_b \Pi^{ab} - \alpha j^a\right] = \nonumber\\
  \left(s -\beta h -\alpha n - \beta w_a q^a \right) \partial_bu^b + 
  \beta w_aq^a\partial_bu^b + 
  \Pi^{ab}\left(\partial_b (\beta u_a) + \partial_a (\beta w_b)\right) - \nonumber\\
  j^a \partial_a \alpha + q^a\partial_a\beta + 
  \beta w^a\left(h\dot u^a+\dot q_a + q^a\partial_b u^b+\partial_b\Pi_a^b) \right)  \nonumber\\
  = \Pi^{ab}\partial_b\beta_a - j^a\partial_a \alpha + 
    q^a \partial_a\beta +\beta w^a \left(\frac{\beta}{h}(h\dot u_a + \dot q_a + q^a\partial_bu^b + partial_b\Pi_a^b)\right) \geq 0 
    \label{epA}
\end{gather}

A further transformation  of the final expression with the help of the momentum balance and also by  Gibbs-Duhem relation \re{GD} results in a remarkable form:
\be
\Sigma =  \left(\Pi^{ab}-w^aq^b\right)\partial_b\beta_a +
    \left( q^a- h w^a\right)(\partial_a\beta -\beta w^b \partial_a u_b)+
    (n w^a-j^a) \partial_a\alpha \geq 0 
\ee{epB}

From this expression it is apparent that the entropy production is zero, whenever the fluid is ideal in the sense of \re{kEM}, suggested by the kinetic theory. The other observation is that 
seen from the thermometer frame $w^a=0$ and the entropy production reduces to
\be
\Sigma =  \Pi^{ab}\partial_b\beta_a + q^a\partial_a\beta - j^a \partial_a\alpha \geq 0 .
\ee{epBthm}
This form of the entropy production does not contain the disturbing acceleration term 
in the thermal thermodynamic force. 
It was suggested to apply  by a different argumentation in \cite{GarSan06a,GarEta09a}. 
However, the generic stability of this fluid is conditional, e.g. $\frac{\partial\beta}{\partial n}|_e >0$ is a requirement. 

The condition of kinetic equilibrium, $q^a=hw^a$, leads to another simplified expression:
\be
\Sigma =  \left(\Pi^{ab}-\frac{q^aq^b}{h}\right)\partial_b\beta_a +
     (\frac{n}{h} q^a-j^a) \partial_a\alpha \geq 0. 
\ee{epC}

In this case only two additive terms occur in the entropy production formula above eq.\re{epC}, 
but for three undetermined constitutive functions $j^q$, $q^a$ and $\Pi^{ab}$. 
Therefore a classical flow-frame fixing, like the Eckart or the Landau-Lifshitz one, 
closes the system of equations instead of restricting or contradicting to thermodynamical requirements. 
In the followings we prove that a rather general choice leads to a generic stable relativistic fluid theory.

\section{Constitutive theory and generic stability in a mixed particle-energy flow-frame}

The fluid velocity in the Eckart flow-frame is given by the direction of the particle number 
density vector: $n^a = N^a/N$, in the Landau-Lifshitz flow-frame by the direction of the 
energy-momentum vector: $e^a = E^a/E$. 
Here $N= \sqrt{N^aN_a}$ and $E= \sqrt{E^aE_a}$ are the magnitudes of the particle number and the energy-momentum density vectors, respectively. Then we introduce a mixed flow, where the coefficients $A$ and $B$ charactarize the weights of the Eckart and Landau-Lifsic flows:
\be
 u^a = \frac{A e^a +B n^a}{O}, \qquad \text{where} \qquad O=\sqrt{A^2 + B^2 + 2AB \chi}.
\ee{mixvel}
 Here $\chi = e^an_a$. The energy and particle number densities in this flow are 
\be
e = u_aE^a = \frac{E}{O}(A+B\chi) \qquad \text{and} \qquad n = u_aN^a = \frac{N}{O}(B+A\chi).
\ee{mdens} 

Moreover, the momentum density and the diffusion current are expressed as
\ba
q^a &=&\frac{E B}{O^2} ((B+A\chi)e^a - (A+B\chi)n^a), \label{q} \\
j^a &=&- \frac{N A}{O^2} ((B+A\chi)e^a - (A+B\chi)n^a).
\ea{j}

By this, these current densities are parallel with a proportionality factor $z$:
\be
j^a = -\frac{NA}{EB} q^a = -zq^a.
\ee{qj_cond}

The entropy production \re{epC} in fact relies on only two constitutive quantities,
$\Pi^{ab}$ and $q^a$:
\be
\Sigma =  \left(\Pi^{ab}-\frac{q^aq^b}{h}\right)\partial_b\beta_a + 
    \left(\frac{n}{h}+z\right)q^a\partial_a\alpha \geq 0 
\ee{epD}

We can identify the corresponding thermodynamic forces and currents as follows:
\begin{center}
\begin{tabular}{c|c|c}
       & Diffusive-Thermal & Viscous  \\ \hline
Fluxes & $\left(\frac{n}{h}+z\right)q^a$ & 
    $\Pi^{ab}-\frac{q^aq^b}{h}$\\ \hline
Forces &  $\nabla^a\alpha $ &
    $\nabla^{(b}\beta^{a)}$
\end{tabular}
\vskip .1cm
Table 1. Thermodynamic fluxes and forces in the particle-energy flow-frame.\end{center}
\vspace{2mm}
Here $\nabla^a = \Delta^{ab}\partial_b$ is the u-space derivative. The corresponding linear response relations for isotropic continua are:
\ba
 \left(\frac{n}{h}+z\right) q^a &=& \lambda \nabla^a\alpha, \label{or_ther}\\
 \Pi^{ab}-\frac{q^aq^b}{h} &=& \zeta\Delta^{ab}\partial^c\beta_c+2\eta \Delta^{\langle bc}\nabla^{a\rangle}\beta_c.
\ea{or_ms}
Here \(\langle \ldots \rangle\) denotes the symmetric traceless part in the bracketed indices, \(\lambda\) is the heat conduction coefficient, \(\zeta\) is the bulk viscosity and \(\eta \) is the shear viscosity.  Because of  the nonnegative entropy production $\lambda\geq 0, \,\,  \zeta \geq 0, \,\, \eta \geq 0$,
reflecting very natural conditions.

Finally the linear stability of the classical homogeneous equilibrium of equations 
\re{pnumc_bal}, \re{e_bal}, \re{momc_bal} with the constitutive equations 
\re{or_ther}-\re{or_ms} has to be discussed. 
The entropy production formula \re{epD} indicates that the equilibrium is characterized 
by zero viscous stress and heat current (momentum density). 
The linearization of the above equations results in:
\ba
 0 &=& \dot{\delta n} +
        n \partial_a \delta u^a - z\partial_a \delta q^a , \label{stab_lnbal}\\
 0 &=& \dot{\delta e} +
        h \partial_a \delta u^a +
        \partial_a \delta q^a, \label{lteb}\\
 0 &=& {h}\dot{\delta u^a}  +
        \Delta^{ab} \partial_b \delta p +
        \dot{\delta q^a} +
        \Delta^a_c \partial_b \delta\Pi^{cb},
        \label{ltib}\\
 0 &=& \left(\frac{n}{h}+z\right)\delta q^a-
        \lambda \Delta^{ab}\partial_b \delta\alpha, \label{lfum}\\
 0 &=&  \delta\Pi^{ab} -
       \zeta \partial_c \delta \beta^c \Delta^{ab} -
        \eta \Delta^{ac}\Delta^{bd}(
      \partial_c\delta \beta_d +\partial_d\delta \beta_c-
            \frac{2}{3}\partial_f\delta \beta^f\Delta_{cd}). 
\ea{lnewm}

One inspects solutions in the form $\delta Q =  Q_0 e^{\Gamma t + i k x}$ 
with a general complex \(\Gamma\). Then the condition of asymptotic stability of the above equations is that the solutions of the characteristic equation \( \det {\bf M} = (\det {\bf N})(\det {\bf R})^2=0\) give  \(Re\ \Gamma< 0\). Here 
\be
  {\bf R} = \begin{pmatrix}
  h\Gamma          & \Gamma     		& ik    & 0\\
  0		   & 1				& 0     & 0\\
  ik\tilde{\eta}   & ik\frac{\tilde{\eta}}{h}   & 1     & 0\\
  ik\tilde{\zeta}  & ik\frac{\tilde{\zeta}}{h} & 0     & 1\\
  \end{pmatrix},
\ee{R}
couples the perturbation fields \(\delta u^a, \delta q^a, \delta \Pi^{xa},\delta \Pi^{aa}\), \(a=y,z\) and 
\be
  {\bf N} = \begin{pmatrix}
  \Gamma & 0   & ikn      & -ik z  & 0 \\
  0   & \Gamma & ikh  & i k  & 0  \\
  ik\partial_n p        & ik\partial_e p  & \Gamma h & \Gamma  & ik \\
  -ik\tilde\lambda\partial_n \alpha & -ik\tilde\lambda\partial_e \alpha  &  0     & 1 & 0 \\
  0   & 0   & ik\tilde{\eta}  & ik\tilde{\eta}/h & 1 
   \end{pmatrix}.
\ee{N}
couples \(\delta n, \delta e, \delta u^x, \delta q^x, \delta \Pi^{xx}\). Here \(h=e+p\),  $\tilde{\eta} = \beta(\zeta +\frac{4}{3}\eta)$,  $\tilde{\zeta} =\beta(\zeta -\frac{2}{3}\eta)$ and \(\tilde \lambda = \lambda (z+n/h)^{-1}\). \(\det {\bf R}=0\) gives the condition 
\be
   h\Gamma + \tilde{\eta} k^2 = 0,
\ee{cond1} 
which results in negative real \(\Gamma\).
\(\det {\bf N}=0 \), on the other hand,  results in a more complicated polinomial
\begin{gather*}
 \Gamma^3 h + 
 \Gamma^2k^2 \left(\tilde{\eta} +\lambda h\partial_n\alpha \right) +\\
 \Gamma k^2 \left(h\partial_e p + n \partial_n p  +
        k^2 \tilde{\eta}\lambda\partial_n \alpha  \right)+  
        k^4\lambda h \left(  
        \partial_e p\partial_n\alpha-\partial_n p \partial_e\alpha\right)= 0.
 \end{gather*}

According to Routh-Hurwitz criteria the real parts of the roots of the polinomial equation $a_0 \Gamma^3+a_1 \Gamma^2 + a_2 \Gamma + a_3 =0$ are always negative if the coefficients are all non-negative and $ a_1 a_2 - a_0 a_3> 0$ \cite{KorKor00b}. It is straightforward to check  that these inequalities are fulfilled, if the transport coefficients \(\lambda>0, \eta>0, \zeta>0\) and the inequalities of the thermodynamic stability are satisfied, 
\be 
\partial_e \beta < 0, \quad 
\partial_n \alpha > 0, \quad 
\partial_e \beta \partial_n \alpha +\left(\partial_n\beta\right)^2 \leq 0.
\ee{conddd2} 
or equivalently $\partial_e^2 s < 0$, $\partial_n^2 s < 0$ and 
$\left(\partial_e\partial_n s\right)^2 \le \partial_e^2 s \cdot \partial_n^2 s$.

\section{Discussion}

Flow-frames are related to thermodynamics and thermodynamics is related to flow-frames. This is inevitable in a relativistic spacetime, where motion separates momentum and energy. Here we have analysed the relation of flow-frames and thermodynamics in case of relativistic fluids, researching the consequences of a motion related thermodynamic framework, where the velocity of the thermometer is not parallel to the flow. 
In this case also the momentum becomes a thermodynamic variable and $w^a$ 
appears as a conjugated intensive variable. It constitutes the flow-orthogonal 
part of the temperature vector: $w^a = T \Delta^{ab}\beta_b$.  

In this case 
\begin{itemize}
 \item trhe conventional thermodynamic forces and currents are interdependent, 
	 viscosity becomes entangled with heat conduction and diffusion.
 \item Kinetic equilibrium is compatible with a preferred frame, where $w^a=q^a/h$. 
 \item A flow-frame fixing selects the proper thermodynamic forces and currents; a dissipative hydrodynamics can be fully determined.
 \item This dissipative hydrodynamics is generic stable in a mixed particle-energy flow-frame, 
	 therefore also in the particular cases of Eckart ($z=0$) and Landau-Lifshitz ($z=\infty$).
\end{itemize}

All previous investigations introduced a flow that was fixed to the thermometer, or, equivalently, a thermodynamics, where the energy density was an extensive variable, but the momentum density was not. It is no wonder, that the Landau-Lifshitz flow-frame appeared to be distinguished for a deeper phenomenological analysis \cite{KosLiu00a,TsuKun13a} and also for the kinetic theory. 

Stability calculations indicate that $w^a$ cannot be arbitrary. In particular the usual thermometer related choice, $w^a=0$, leads to violently unstable theories \cite{HisLin83a}, when the the additionally introduced flow-frame is incompatible, say the Eckart one. 

The general form of the entropy production \re{epB} shows that the choice of a flow-frame is not 
merely a possibility, but it is a necessity for fixing the consitutive relations with all its implications
to the described physical system.

\section{Acknowledgement}

The work was supported by the grants Otka K81161 and K104260. The authors thank to Tam\'as Matolcsi and Etele Moln\'ar for valuable discussions.


\begin{thebibliography}{10}

\bibitem{Eck40a3}
Carl Eckart.
\newblock The thermodynamics of irreversible processes, {III}. {R}elativistic
  theory of the simple fluid.
\newblock {\em Physical Review}, 58:919--924, 1940.

\bibitem{LanLif59b}
L.~D. Landau and E.~M. Lifshitz.
\newblock {\em Fluid mechanics}.
\newblock Pergamon Press, London, 1959.

\bibitem{GroAta80b}
S.R. de~Groot, W.~A. van Leeuwen, and Ch.~G. van Weert.
\newblock {\em Relativistic Kinetic Theory}.
\newblock North Holland, Amsterdam, 1980.

\bibitem{Mat05b}
T.~Matolcsi.
\newblock {\em Ordinary thermodynamics}.
\newblock Akad\'emiai Kiad\'o (Publishing House of the Hungarian Academy of
  Sciences), Budapest, 2005.

\bibitem{HisLin83a}
W.~A. Hiscock and L.~Lindblom.
\newblock Stability and causality in dissipative relativistic fluids.
\newblock {\em Annals of Physics}, 151:466--496, 1983.

\bibitem{HisLin87a}
W.~A. Hiscock and L.~Lindblom.
\newblock Linear plane waves in dissipative relativistic fluids.
\newblock {\em Physical Review D}, 35(12):3723--3731, 1987.

\bibitem{VanBir08a}
P.~V\'an and T.~S. B\'\i{}r\'o.
\newblock Relativistic hydrodynamics - causality and stability.
\newblock {\em The European Physical Journal - Special Topics}, 155:201--212,
  2008.
\newblock arXiv:0704.2039v2.

\bibitem{Van09a}
P.~V\'an.
\newblock Generic stability of dissipative non-relativistic and relativistic
  fluids.
\newblock {\em Journal of Statistical Mechanics: Theory and Experiment}, page
  02054, 2009.
\newblock arXiv: 0811.0257.

\bibitem{BirVan10a}
T.~S. B\'ir\'o and P.~V\'an.
\newblock About the temperature of moving bodies.
\newblock {\em EPL}, 89:30001, 2010.
\newblock arXiv:0905.1650v1.

\bibitem{VanBir12a}
P.~V\'an and T.S. Bir\'o.
\newblock First order and generic stable relativistic dissipative
  hydrodynamics.
\newblock {\em Physics Letters B}, 709(1-2):106--110, 2012.
\newblock arXiv:1109.0985[nucl-th].

\bibitem{VanBir13p}
P.~V\'an and T.~Bir\'o.
\newblock Dissipation flow-frames: particle, energy, thermometer.
\newblock In M.~Pilotelli and G.~P. Beretta, editors, {\em Proceedings of the
  12th Joint European Thermodynamics Conference}, pages 546--551, Brescia,
  2013. Cartolibreria SNOOPY.
\newblock arXiv:1305.3190.

\bibitem{HisLin85a}
W.~A. Hiscock and L.~Lindblom.
\newblock Generic instabilities in first-order dissipative relativistic fluid
  theories.
\newblock {\em Physical Review D}, 31(4):725--733, 1985.

\bibitem{Van08a}
P.~V\'an.
\newblock Internal energy in dissipative relativistic fluids.
\newblock {\em Journal of Mechanics of Materials and Structures},
  3(6):1161--1169, 2008.
\newblock arXiv:07121437 [nucl-th].

\bibitem{Van11p}
P.~V\'an.
\newblock Kinetic equilibrium and relativistic thermodynamics.
\newblock {\em EPJ WEB of Conferences}, 13:07004, 2011.
\newblock arXiv:1102.0323.

\bibitem{Mul69a}
I.~M\"uller.
\newblock Toward relativistic thermodynamics.
\newblock {\em Archive for Rational Mechanics and Analysis}, 34(4):259--282,
  1969.

\bibitem{DenEta08a}
G.~S. Denicol, T.~Kodama, T.~Koide, and P.H. Mota.
\newblock Stability and causality in relativistic dissipative hydrodynamics.
\newblock {\em Journal of Physics G}, 35:115102, 2008.

\bibitem{PuEta10a}
Tomoi~Koide Shi~Pu and Dirk~H. Rischke.
\newblock Does stability of relativistic dissipative fluid dynamics imply
  causality?
\newblock {\em Physical Review D}, 81:114039, 2010.

\bibitem{Fic92a}
G.~Fichera.
\newblock Is the {F}ourier theory of heat propagation paradoxical?
\newblock {\em Rediconti del Circolo Matematico di Palermo}, XLI:5--28, 1992.

\bibitem{KosLiu00a}
P.~Kost\"adt and M.~Liu.
\newblock On the causality and stability of the relativistic diffusion
  equation.
\newblock {\em Physical Reviews D}, 62:023003, 2000.

\bibitem{Cim04a}
V.~A. Cimmelli.
\newblock On the causality requirement for diffusive-hyperbolic systems in
  non-equilibrium thermodynamics.
\newblock {\em Journal of Non-Equilibrium Thermodynamics}, 29(2):125--139,
  2004.

\bibitem{LiuAta86a}
I-S. Liu, I.~M\"uller, and T.~Ruggeri.
\newblock Relativistic thermodynamics of gases.
\newblock {\em Annals of Physics}, 169:191--219, 1986.

\bibitem{PerCal10a}
J.~Peralta-Ramos and E.~Calzetta.
\newblock Divergence-type nonlinear conformal hydrodynamics.
\newblock {\em Physical Review D}, 80:126002, 2010.

\bibitem{Ger01m}
R.~Geroch.
\newblock On hyperbolic "theories" of relativistic dissipative fluids.
\newblock 2001.
\newblock arXiv:gr-qc/0103112.

\bibitem{GerLin90a}
R.~Geroch and L.~Lindblom.
\newblock Dissipative relativistic fluid theories of divergence type.
\newblock {\em Physical Review D}, 41:1855--1861, 1990.

\bibitem{Mol09a}
E.~Moln\'ar.
\newblock Comparing the first and second order theories of relativistic
  dissipative fluid dynamics using the 1+1 dimensional relativistic flux
  corrected transport algorithm.
\newblock {\em European Physical Journal C}, 60:413--429, 2009.

\bibitem{BouEta10a}
I.~Bouras, E.~Moln\'ar, H.~Niemi, Z.~Xu, A.~El, O.~Fochler, C.~Greiner, and
  D.~H. Rischke.
\newblock Investigation of shock waves in the relativistic riemann problem: A
  comparison of viscous fluid dynamics to kinetic theory.
\newblock {\em Physical Review C}, 82:024910, 2010.

\bibitem{Isr63a}
W.~Israel.
\newblock Relativistic kinetic theory of a simple gas.
\newblock {\em Journal of Mathematical Physics}, 4(9):1163--1181, 1963.

\bibitem{Far894a}
Gy. Farkas.
\newblock A {F}ourier-f\'ele mechanikai elv alkalmaz\'asai.
\newblock {\em Mathematikai \'es Term\'eszettudom\'anyi \'Ertes\'\i{}t\H{o}},
  12:457--472, 1894.
\newblock in Hungarian.

\bibitem{Min896b}
H.~Minkowski.
\newblock {\em Geometrie der Zahlen}.
\newblock Teubner, Leipzig und Berlin, 1896.

\bibitem{Sch98b}
A.~Schriver.
\newblock {\em Theory of linear and integer programming}.
\newblock John Wiley and Sons, Chicester-etc.., 1998.

\bibitem{TsuKun12a}
K.~Tsumura and T.~Kunihiro.
\newblock Derivation of relativistic hydrodynamic equations consistent with
  relativistic {B}oltzmann equation by renormalization-group method.
\newblock {\em The European Physical Journal A}, 48:162, 2012.

\bibitem{GarSan06a}
L.~S. Garc\'\i{}a-Col\'\i{}n and A.~Sandoval-Villalbazo.
\newblock Relativistic non-equilibrium thermodynamics revisited.
\newblock {\em Journal of Non-Equilibrium Thermodynamics}, 31:11--22, 2006.

\bibitem{GarEta09a}
A.~L. Garcia-Perciante, L.~S. Garcia-Colin, and A.~Sandoval-Villalbazo.
\newblock On the nature of the so-called generic instabilities in dissipative
  relativistic hydrodynamics.
\newblock {\em General Relativity and Gravitation}, 41(7):1645--1654, 2009.

\bibitem{KorKor00b}
G.~A. Korn and T.~M. Korn.
\newblock {\em Mathematical Handbook for Scientists and Engineers: Definitions,
  Theorems, and Formulas for Reference and Review}.
\newblock Dover, 2nd, revised edition, 2000.

\bibitem{TsuKun13a}
K.~Tsumura and T.~Kunihiro.
\newblock Uniqueness of {L}andau-{L}ifshitz energy frame in relativistic
  dissipative hydrodynamics.
\newblock {\em Physical Review E}, 87:053008, 2013.

\end{thebibliography}

\end{document}